\documentclass[10pt,conference,a4paper]{IEEEtran}

\usepackage{amsmath,amssymb}
\usepackage{cite}
\usepackage{algorithm}
\usepackage{algpseudocode}
\usepackage{booktabs}
\usepackage{tabularx}
\usepackage{array}
\usepackage{graphicx}
\usepackage{xcolor}
\usepackage{microtype}
\usepackage{url}
\usepackage{enumitem}
\usepackage{flafter}
\usepackage{placeins}
\usepackage{balance}
\usepackage{hyperref}

\newcolumntype{Y}{>{\raggedright\arraybackslash}X}
\newcommand{\method}{RouteBR}
\title{Route Before Response: Fail-Closed Routing for\\
LLM-Enabled Customer-Service Software}

\author{
\IEEEauthorblockN{Yuchen Cao\textsuperscript{1}, Zhenyu Mao\textsuperscript{1}, Jacky Wai Keung\textsuperscript{1},
Yifei Wang\textsuperscript{1}, Guancheng Lin\textsuperscript{1}, Jingyu Zhang\textsuperscript{2,*}}
\IEEEauthorblockA{\textsuperscript{1}City University of Hong Kong, Hong Kong\\
\{cynthcao2-c, zhenyumao2-c, ywang4748-c, guanchlin4-c\}@my.cityu.edu.hk,
Jacky.Keung@cityu.edu.hk}
\IEEEauthorblockA{\textsuperscript{2}Hong Kong Metropolitan University, Hong Kong\\
\textsuperscript{*}Corresponding author: fzhang@hkmu.edu.hk}
}

\begin{document}
\maketitle

\begin{abstract}
LLM-enabled customer-service software in financial institutions must choose a downstream service before responding. Classifiers and LLM routers can identify a relevant label or destination, but a composed application must still verify dispatch permission and entity registration. We study this integration problem as \emph{boundary-aware pre-response routing}. We present \method, a controller in which one model call proposes a primary service action and registered object, then deterministic code checks identifiers and versioned rules before dispatch or a specified non-dispatch outcome. Lower-ranked candidates cannot replace a blocked primary route. We construct 600 privacy-preserving scenarios from recurring confidential-request patterns at a real financial institution; three engineers jointly reviewed the scenarios and labels. On the 200-case CONFIRM MAIN split, \method{} reaches 97.5\% final-path exact match. A matched one-call baseline given the same rules reaches 95.5\%, whereas a diagnostic baseline without the rule program obtains 94.0\% action accuracy but only 41.0\% final-path exact match. Across 166 non-dispatch MAIN and FLIP cases, \method{} produces no false dispatch. The artifact pairs all 1,960 prediction rows with raw receipts while excluding gold and reviewer fields from model requests. These results support separating model proposals from code-owned, replayable execution. Our implementation is available at {\footnotesize\href{https://github.com/YgcCoder/RouteBR}{\mbox{\texttt{https://github.com/YgcCoder/RouteBR}}}}.
\end{abstract}

\begin{IEEEkeywords}
large language models, LLM-enabled software, customer service, financial software, service routing
\end{IEEEkeywords}

\section{Introduction}

LLM-enabled customer-service software in financial institutions increasingly connects one conversational entry point to information lookup, account analysis, advisory services, support workflows, and protected operations. Tool-augmented language models can interpret open-ended requests and construct calls to external services~\cite{schick2023toolformer,li2023api,qin2024toolllm,yao2022react}. These destinations expose distinct interfaces, identifiers, and runtime preconditions. Together, they make the entry point a software integration boundary that passes requests, identifiers, and responsibility to downstream components.

\begin{figure}[!t]
\centering
\includegraphics[width=\columnwidth]{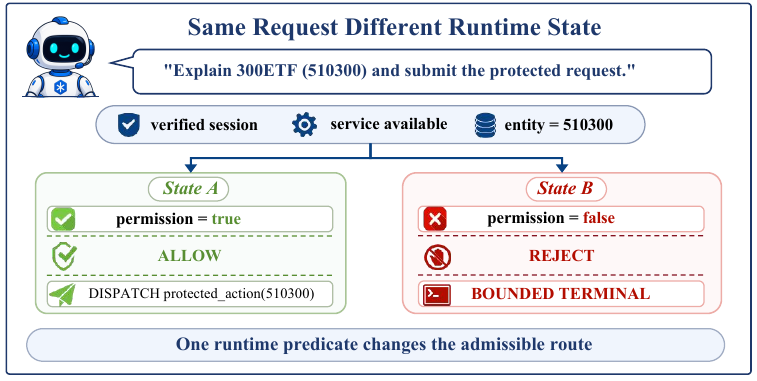}
\caption{Motivating state-conditioned route. The request and object are fixed; changing only \texttt{permission} from true to false changes the result from protected-operation dispatch to rejection. Matching the request to a service does not by itself determine an executable route.}
\label{fig:motivation}
\end{figure}

Yet a service that matches the request is not necessarily allowed by the current software state and policy, and a referenced entity may not map to a valid system identifier~\cite{wu2020scalable,rebedea2023nemo,yang2025adaptiveguard,wang2025agentspec}. Figure~\ref{fig:motivation} fixes the request, history, and referenced object while changing only one permission value. One record dispatches the protected operation; the otherwise identical record ends in rejection. The correct downstream route therefore depends on runtime information that is absent from the request text.

A routing error at this boundary changes executable software behavior. For this study, a route has three fields: an \emph{action}, an \emph{object}, and an \emph{outcome}. The action is one of five service operations: advisory, information lookup, account analysis, support, or a protected operation. When required, the object is a registered identifier used by the service; in our benchmark, it is usually a canonical ETF identifier. The outcome is either \textsc{dispatch} or one of four non-dispatch results: \textsc{clarify}, \textsc{defer}, \textsc{handoff}, or \textsc{reject}. Together, these fields define the final route. A false dispatch sends a request that should stop; a false block stops one that should be dispatched. Even valid JSON may name an unknown object or request an operation forbidden by the current state.

These failures define \emph{boundary-aware pre-response routing}. Its input combines a request and history with a service catalog, registered objects, runtime state, and versioned rules. Its software-visible output is either an allowed action--object dispatch or a specified non-dispatch outcome. Unlike destination selection alone, this task requires the semantic action, referenced object, current state, and final outcome to agree before execution. The objective is to preserve flexible language interpretation while ensuring that the selected service is permitted, every required object is registered, and the decision can be replayed from recorded software inputs. The acceptance object is therefore a complete executable route rather than an isolated label.

Prior work covers individual stages. Intent and out-of-scope methods return labels or abstain, LLM routers choose destinations, and tool-use methods produce calls and arguments~\cite{larson2019evaluation,hendrickx2024machine,ong2025routellm,patil2024gorilla}. Entity linking maps mentions to registered identifiers, and runtime controls constrain later transitions~\cite{wu2020scalable,rebedea2023nemo,yang2025adaptiveguard,wang2025agentspec}. Without a separate pre-dispatch verification layer, these outputs do not establish a complete route in our setting. Three conditions remain: the service matches the request, state and policy permit it, and any required object resolves to a registered identifier. A local result can appear correct while the composed system dispatches a forbidden service or invalid object~\cite{ruan2024identifying,debenedetti2024agentdojo,zhan2024injecagent,xue2025characterization}; evidence from source-code tasks likewise cautions against relying on an LLM's own verification of flawed outputs~\cite{yu2024fight}.

We present \textbf{\method}, which stands for \textbf{Route} \textbf{B}efore \textbf{R}esponse, as a pre-response controller for this boundary. One model call proposes a primary service action and registered object. The controller keeps that primary choice, so a lower-ranked alternative cannot replace it after a rule blocks dispatch. Deterministic code applies the recorded state and policy, and the object check verifies the registered identifier before the controller emits either a typed dispatch or a specified non-dispatch outcome. Alternatives are recorded for inspection only and cannot be dispatched. In this way, the model interprets the request while ordinary code owns the final software transition. Our contributions are:

\begin{itemize}[leftmargin=*,nosep]
  \item We formulate boundary-aware pre-response routing as a software decision over a service action, registered object, runtime state, and final routing outcome.
  \item We design \method, which separates model proposals from code-owned state checks, closed-list object validation, controlled repair, and replayable decision traces.
  \item We construct a privacy-preserving benchmark of 600 synthetic reconstructions from recurring confidential-request patterns, jointly reviewed by three engineers from the institution and organized into MAIN, FLIP, and ENTITY cases.
  \item We evaluate complete-path correctness, state-conditioned execution, the closed-list object contract, and backend portability through matched one-call controls, two controlled mechanism checks, and six hosted model backends.
\end{itemize}

\section{Related Work}

\subsection{Customer-Service Task Formulations}

Financial question-answering benchmarks evaluate answers over documents and tables~\cite{chen2021finqa,islam2023financebench}. Task-oriented dialogue represents services through intents, slots, state, and API arguments~\cite{eric2020multiwoz,rastogi2020towards,byrne2019taskmaster}; intent, out-of-scope, and selective prediction add labels or abstention~\cite{casanueva2020efficient,larson2019evaluation,geifman2017selective,hendrickx2024machine}. These tasks motivate service descriptions, history, and bounded refusal, but their target is principally an answer, label, or call inferred from dialogue. Our boundary instead receives runtime state as a separately versioned software record. This permits paired tests in which request-side input is fixed while one predicate and the admissible output change.

\subsection{Semantic Routing and Tool Calling}

Routing research evaluates different output objects. Autoregressive semantic route planning predicts navigation paths from natural-language criteria~\cite{zhao2024semantic}, whereas LLM routers select models or services and guarded variants add rejection or prompt-optimized routing behavior~\cite{ong2025routellm,vsleher2025guarded,vsleher2026influence}. Tool-use benchmarks assess interface choice, arguments, execution, or API collections~\cite{li2023api,qin2024toolllm,zhuang2023toolqa,patil2024gorilla}, while $\tau$-bench combines interaction, tools, policies, and final database state~\cite{yao2024tau}. These studies establish the relevance of routing and policy following. \method{} narrows the evaluated output to one pre-dispatch action--object pair or a specified non-dispatch outcome. A wrong outcome is a route error, dispatch when the gold route should stop is a safety error, and refusal under \textsc{allow} is an availability error. Table~\ref{tab:positioning} compares these evaluated outputs, not whether the other systems could be extended.

The distinction also changes the oracle. Router evaluation commonly asks whether the selected destination is appropriate, while tool-use evaluation may continue through arguments, execution, and task outcome. Our controller ends earlier but requires a stronger precondition at that point: the selected action, canonical object, versioned state, and terminal must agree before any response service is invoked. A correct bounded terminal cannot erase a wrong semantic action from diagnosis, and a correct action cannot compensate for dispatch under a prohibiting state. This contract-level oracle motivates reporting path, false-dispatch, and false-block measures together rather than treating routing as destination accuracy alone.

\subsection{Entity Linking and Runtime Control}

Named-entity recognition identifies spans and types; entity linking maps mentions to canonical identifiers~\cite{tedeschi2022multinerd,wu2020scalable,de2020autoregressive,de2022multilingual,botha2020entity}. \method{} uses a smaller closed-list problem: the supplied registry defines the only dispatchable objects, and ambiguity becomes \textsc{unresolved}. Runtime rails, adaptive guards, and executable specifications separately constrain model-driven transitions~\cite{rebedea2023nemo,yang2025adaptiveguard,wang2025agentspec}. \method{} joins these restrictions after semantic selection: code checks the committed action against state and its object against the catalog before invocation. A trace can therefore attribute a bounded path to a named rule, unresolved object, or persistent interface failure rather than to a generic refusal. This scoped contract complements general NER and agent-safety work with behavior that is directly testable at a software dispatch boundary.

Closed-list linking serves here as an executable interface contract. The registry is supplied for the current request, and the relevant question is whether a proposed identifier is both canonical and permitted by the selected action contract. An out-of-registry string is therefore more than a linking error: if allowed to cross the boundary, it becomes an invalid downstream argument. Conversely, resolving every ambiguity by refusal can be safe but unavailable. This connection between canonical identity and dispatchability is why RQ3 evaluates object accuracy together with final path and false block.

\begin{table}[!t]
\caption{Representative formulations and evaluated properties. Obj.: closed object; State: runtime predicates; Typed: action contract; Code: deterministic transition owner; Flip: paired state test.}
\label{tab:positioning}
\centering
\footnotesize
\setlength{\tabcolsep}{1.2pt}
\renewcommand{\arraystretch}{1.27}
\begin{tabularx}{\columnwidth}{@{}Y>{\raggedright\arraybackslash}p{0.17\columnwidth}ccccc@{}}
\toprule
Formulation & Output & Obj. & State & Typed & Code & Flip \\
\midrule
Intent/OOS~\cite{casanueva2020efficient,larson2019evaluation} & Label/reject & $\times$ & $\times$ & $\times$ & $\times$ & $\times$ \\
Entity linking~\cite{wu2020scalable,de2020autoregressive} & Canonical ID & \checkmark & $\times$ & $\times$ & $\times$ & $\times$ \\
LLM routing~\cite{ong2025routellm} & Destination & $\times$ & $\times$ & $\times$ & $\times$ & $\times$ \\
Guarded routing~\cite{vsleher2025guarded} & Service/reject & $\times$ & $\times$ & $\times$ & $\times$ & $\times$ \\
Tool use~\cite{li2023api,qin2024toolllm,patil2024gorilla} & Tool call & $\times$ & $\times$ & \checkmark & $\times$ & $\times$ \\
Runtime control~\cite{wang2025agentspec} & Transition & $\times$ & \checkmark & \checkmark & \checkmark & $\times$ \\
\textbf{\method(ours)} & Final path & \checkmark & \checkmark & \checkmark & \checkmark & \checkmark \\
\bottomrule
\end{tabularx}
\end{table}

Table~\ref{tab:positioning} compares the reported evaluation objects of each formulation. A check means that the cited formulation explicitly evaluates the property; a cross means that the property lies outside its reported evaluation target. Boundary-aware pre-response routing combines the closed object, versioned state, typed action contract, deterministic transition owner, and paired state test in one pre-dispatch decision.

\FloatBarrier
The comparison identifies the checks required at component handoffs. Dialogue models supply request-side structure, routers supply candidate destinations, entity linkers supply registered identifiers, and runtime-control work supplies executable guards. Boundary-aware pre-response routing evaluates how these outputs compose: the semantic stage commits to one current action, the object remains inside a supplied registry, and code reproduces the state transition from recorded inputs. Its regression oracle must therefore preserve the relation among action, object, state, and outcome. Testing only the action can accept an unsafe dispatch, while testing only the outcome can conceal a wrong but conservatively blocked route.

\section{\method}
\label{sec:routebr}

\subsection{Overview and Formulation}

\begin{figure*}[!t]
\centering
\includegraphics[width=\textwidth]{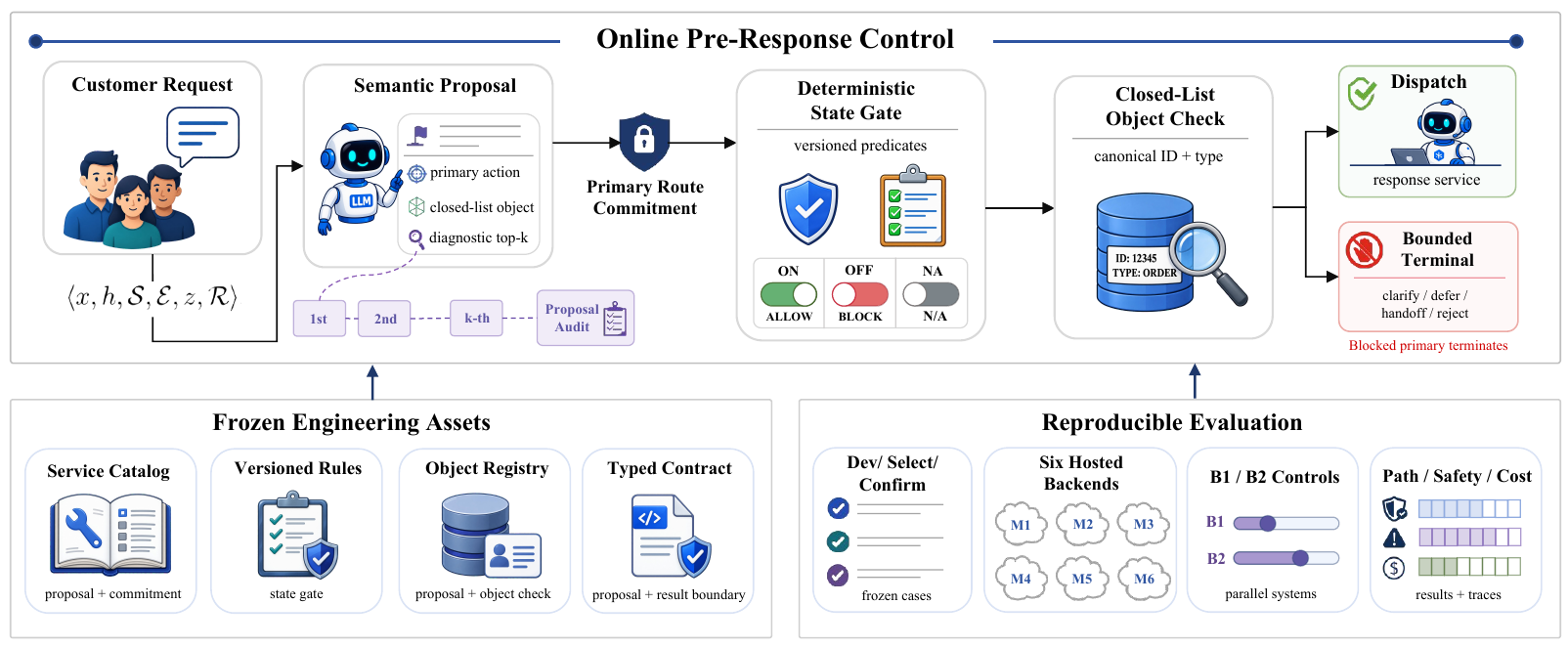}
\caption{\method{} execution and evaluation framework. The online path commits to one semantic proposal, then code applies state and object checks before dispatch or a specified non-dispatch outcome; lower-ranked candidates are recorded only for inspection and cannot be dispatched. Frozen assets feed their corresponding control points, not the customer-request card. Evaluation wraps and measures the controller rather than entering its online path.}
\label{fig:overview}
\end{figure*}

Figure~\ref{fig:overview} places \method{} between a customer entry point and downstream response services. Its input is $I=\langle x,h,\mathcal{S},\mathcal{E},z,\mathcal{R}\rangle$: request $x$, bounded history $h$, versioned service catalog $\mathcal{S}$, closed entity registry $\mathcal{E}$, runtime state $z$, and prioritized rules $\mathcal{R}$. The output is either a dispatch record $(a,o)$ or one terminal in $\{\textsc{clarify},\textsc{defer},\textsc{handoff},\textsc{reject}\}$. Customer-facing response generation is outside this boundary.

The trust boundary separates three owners. Product engineers own the catalog and the versioned rule set; the session and service infrastructure own $z$; the model contributes only a typed semantic proposal. A downstream response service is not addressable from the model adapter. It can be reached only through the controller's dispatch return, after all contracts have succeeded. Consequently, a parseable model output is evidence for semantic selection, not authorization. This division also makes configuration changes local: a new rule version can alter admissibility without changing the prompt, whereas a catalog change can be reviewed as an interface migration.

The lower panels represent offline dependencies and evaluation. Frozen assets map to distinct control points: the catalog constrains proposal and action validation; rules drive the state gate; the registry supplies candidates and the closed-list check; and the typed contract validates proposal and result records. The catalog has five action families with stable identifiers and permitted object types; eight synthetic predicates and prioritized rules with a default \textsc{allow} model the boundary without reproducing institutional policy. Evaluation wraps the controller: DEV, SELECT, and CONFIRM supply cases, backends replace only the proposer, B1 and B2 run in parallel, and path, safety, and cost metrics consume final results and traces.

Table~\ref{tab:catalog} summarizes the interface surface. The names describe response-service families, not customer-visible business units. Four actions are object-sensitive, while support can be object-free. This difference is encoded in the catalog rather than inferred after routing. A protected action is not special-cased in the model prompt; its additional identity and permission requirements are ordinary policy rules evaluated only after it becomes the committed primary action.

\begin{table}[!b]
\caption{Closed action contracts used by \method. All identifiers are versioned catalog values.}
\label{tab:catalog}
\centering
\footnotesize
\setlength{\tabcolsep}{2.4pt}
\renewcommand{\arraystretch}{1.13}
\begin{tabular}{@{}lll@{}}
\toprule
ID & Response-service family & Permitted object type \\
\midrule
A1 & Advisory & ETF \\
A2 & Information lookup & ETF or topic \\
A3 & Account analysis & Account or ETF \\
A4 & Support & None or service \\
A5 & Protected action & ETF or account \\
\bottomrule
\end{tabular}
\end{table}

The model produces $q=\langle a,o,K\rangle$, where $a\in\mathcal{S}$ is the primary action, $o\in\mathcal{E}\cup\{\textsc{none},\textsc{unresolved}\}$, and $K$ contains at most three unique catalog actions. The contract requires $a=K_1$. This equality turns ranking into an executable commitment: $K_2$ and $K_3$ are retained for audit, but never become fallback routes after policy evaluation.

We define a final path as the pair of an executable action and object when the terminal is \textsc{dispatch}, or as one bounded terminal otherwise. Exact-match evaluation therefore compares the software-visible result, not an internal ranking. Two proposals with the same action can differ in final path because their objects differ; two identical proposals can also produce different final paths because $z$ selects different policy outcomes. This definition is what links semantic routing to the integration contract exercised by the experiments.

Figure~\ref{fig:motivation} illustrates the resulting composition. Both records provide the same request, history, catalog entry, and canonical ETF. The proposal contract should therefore produce the same protected-action pair. The state gate then reads the only changed field, \texttt{permission}, and selects either the named \textsc{allow} rule or the bounded \textsc{reject} rule. Only the former reaches the object contract and dispatch return. This controlled pair holds semantic evidence fixed while one versioned runtime fact changes the expected path.

This decomposition yields three independently testable interfaces. The proposal interface accepts only catalog and registry identifiers. The enforcement interface accepts an already validated primary action plus a state and rule version, and returns a named boundary decision. The dispatch interface accepts only an \textsc{allow} result and a validated action--object pair. A release can therefore replay policy and object checks from a stored proposal without calling the hosted model. Conversely, prompt or backend changes can be tested against frozen state and policy assets. The implementation and evaluation use these seams to separate semantic, state, object-validation, and adapter errors rather than assigning every wrong path to the model.

\subsection{Proposal and Primary Commitment}

The prompt receives request text, dialogue history, catalog descriptions, and candidate entities, but not runtime predicates or rules. It selects the current requested function rather than a hypothetical later action, and resolves an object only through exact code, name, alias, or unique antecedent evidence. The prompt returns one JSON object with \texttt{action\_id}, \texttt{object\_id}, and \texttt{topk}. This separation prevents the semantic call from adapting its preferred action to whatever runtime route happens to be allowed.

The prompt uses the catalog and entity arrays as data, not instructions. It gives priority to a complete current request with an explicit object or imperative expression; emotional language, conditional future operations, and supporting explanations remain secondary cues. For object resolution, exact codes and aliases precede display names and a unique antecedent in the bounded history. Similarity matching and generated identifiers are forbidden. These constraints make every accepted value checkable against the supplied interface, while action and final-path metrics evaluate semantic correctness separately.

The validator rejects unknown or duplicated actions, disagreement between the primary action and $K_1$, out-of-registry objects, and object types incompatible with the primary action. It permits one format-only repair that receives the invalid output and validation errors but may not replace semantic evidence or closed identifiers. A second failure reaches a configured non-dispatch terminal. The accepted proposal and every validation error remain in the trace.

Validation proceeds from syntax to enumerated identifiers and then to cross-field relations. This order separates a malformed provider response from an unknown catalog value, and separates both from an object whose type is incompatible with an otherwise valid action. The trace retains the first failed contract and the supplied value rather than only the eventual terminal. As a result, fail-closed behavior does not erase the maintenance signal: adapter owners can investigate schema failures, catalog owners can inspect action errors, and registry owners can review unresolved or mistyped objects without inferring the cause from a generic refusal.

Repair is deliberately narrower than a second routing attempt. The original payload, invalid output, and error list are preserved, and the repair instruction asks only for a contract-conforming representation of the same semantic choice. This prevents repeated calls from searching for an action that happens to pass the policy. If the corrected object still lies outside $\mathcal{E}$, if the primary action changes, or if the output again violates the schema, the controller terminates. The experiments count repair as an additional API call and report first-pass JSON validity separately.

Primary commitment is the key compositional invariant. Suppose the request supports a protected action and an advisory alternative. If the protected action is primary but current permission is false, the controller returns the rule's terminal. It does not reinterpret the lower-ranked advisory candidate as an executable substitute. This prevents semantic ranking from becoming an implicit policy-bypass mechanism.

This invariant also clarifies responsibility for mixed-service language. The model must choose which function is current and primary; the policy engine decides only whether that action is admissible. The engine does not rank semantics, and the model does not search the policy space. A wrong primary action remains a measurable semantic error rather than being concealed by a later fallback. Diagnostic alternatives can still help engineers inspect ambiguity or revise catalog descriptions, but they have no execution authority.

\subsection{Deterministic Enforcement and Object Validation}

For the committed action $a$, the policy engine selects the first matching rule:
\begin{equation}
b(a,z,\mathcal{R})=r_j.\mathrm{decision},\quad
j=\arg\min_{i:r_i(a,z)=\mathrm{true}} r_i.\mathrm{priority}.
\end{equation}
The model does not predict $b$ and cannot override it. If $b\neq\textsc{allow}$, \method{} records the winning rule and returns the corresponding terminal. The state-removal intervention in RQ2 bypasses this transition while keeping the same semantic proposal, exposing the errors prevented by code-owned policy.

Rules are ordered by numeric priority and then stable rule identifier. When several rules match, the rule with the smallest priority number determines the routing outcome. In the evaluated cases, the active non-default predicates cover safety escalation, unsupported requests, service unavailability, unclear scope, unresolved entity state, unverified identity, and missing protected-operation permission. A final empty-condition rule returns \textsc{allow}, making the policy function total for every catalog action. This explicit default is important for replay: absence of a matching prohibition is represented by a named rule rather than by missing control flow.

When $b=\textsc{allow}$, the object contract checks whether $a$ requires an object, whether $o$ belongs to the supplied registry, and whether its type is permitted by the catalog. An unresolved required object returns \textsc{clarify}; object-free support paths require \textsc{none}. The closed-resolver intervention in RQ3 makes object-sensitive proposals unresolved. It measures the availability cost of losing closed-list object validation rather than silently dispatching an unchecked object.

Let $b=b(a,z,\mathcal{R})$ and let $\bot_o$ denote an unresolved object. The final transition is
\begin{equation}
T(q,z)=
\begin{cases}
b, & b\neq\textsc{allow},\\
\textsc{clarify}, & b=\textsc{allow},\ o=\bot_o,\\
\textsc{dispatch}(a,o), & b=\textsc{allow},\ C_{a}(o),
\end{cases}
\end{equation}
where $C_a$ is the action-specific closed-object contract. Persistent validation failure is handled before $T$ and reaches the configured fail-closed terminal. The reference implementation enforces three resulting invariants: every dispatch action belongs to $\mathcal{S}$; every required dispatch object belongs to $\mathcal{E}$ and satisfies $C_a$; and no dispatch is returned when the winning rule is non-\textsc{allow}. These are construction properties of the control path, while semantic correctness of $a$ and $o$ remains empirical.

State pairing gives a direct test of this separation. For two inputs that differ only in $z$, the semantic proposal should remain unchanged. The boundary may change because $b$ is a pure function of the committed action, state record, and rule version. Conversely, if a model changes $a$ for identical language across a pair, the error is visible rather than absorbed into the rule engine. RQ2 requires both pair members to reach their expected final paths.

\begin{algorithm}[!t]
\caption{Route before response}
\label{alg:routebr}
\footnotesize
\begin{algorithmic}[1]
\Require request $x$, history $h$, catalog $\mathcal{S}$, entities $\mathcal{E}$, state $z$, rules $\mathcal{R}$
\Ensure typed dispatch $(a,o)$ or bounded terminal
\State $q\gets\Call{Propose}{x,h,\mathcal{S},\mathcal{E}}$
\State $q\gets\Call{ValidateOrRepair}{q}$
\If{$q=\bot$} \State \Return \Call{FailClosed}{} \EndIf
\State $(a,o,K)\gets q$; \textbf{assert} $a=K_1$
\State $b\gets\Call{FirstMatchingRule}{a,z,\mathcal{R}}$
\If{$b\neq\textsc{allow}$} \State \Return $b$ \EndIf
\If{$a$ requires object and $o=\textsc{unresolved}$} \State \Return \textsc{clarify} \EndIf
\State \textbf{assert} $a\in\mathcal{S}$ and $o$ satisfies the closed object contract
\State \Return \Call{Dispatch}{$a,o$}
\end{algorithmic}
\end{algorithm}

\subsection{Executable Contract and Implementation}

Algorithm~\ref{alg:routebr} summarizes the complete control path. The reference implementation defines typed records for requests, services, entities, rules, traces, and results; a replay backend for tests; an OpenAI-compatible adapter; and a deterministic policy engine. Ten unit tests cover normal dispatch, invalid JSON, out-of-list objects, primary-route commitment, state removal, object-validation removal, and adapter behavior. These tests establish control-flow properties, not empirical routing accuracy.

Configuration loading follows the same fail-closed rule as request handling. Controller construction rejects an empty catalog and rejects \textsc{dispatch} as a fallback terminal. The policy engine requires at least one rule, rejects duplicate rule identifiers, and raises an error if evaluation reaches the end without an explicit default. Rules are sorted by numeric priority and stable identifier before any request is processed. These checks move configuration defects to startup or replay instead of converting them into an implicit runtime choice. They also make a policy snapshot self-contained: given an accepted proposal, state record, and rule version, the winning boundary is deterministic.

The unit tests exercise each executable branch with a replay backend. They verify an ordinary dispatch, a blocked primary action whose allowed secondary remains diagnostic, an unresolved object, persistent out-of-list output after repair, the two safeguard removals, and an object-free service. Adapter tests separately check response identity and receipt handling. Each assertion targets the public result and trace rather than a private model explanation. The benchmark then tests the part these unit tests cannot establish: whether the hosted model chooses the correct semantic action and object on varied customer-service language. This division keeps construction guarantees, integration tests, and empirical accuracy as distinct evidence.

The implementation is dependency-free Python at the controller layer. Immutable catalog, entity, request, and rule records make version changes explicit, while the result object contains terminal, action, object, reason codes, and a trace. Policy evaluation is a stable linear scan over the prioritized rules; object membership is a hash lookup over the request's candidate registry. With the small catalog used here, deterministic work is negligible relative to the model call, but the separation also avoids sending policy reasoning through multiple model turns.

The implementation records requested and response-reported model identities, prompt and input hashes, token usage, latency, repair status, and the accepted transition. Credentials remain outside the run ledger. The controller makes one semantic call per case unless bounded repair is needed, so its call count is directly comparable with the two one-call baselines.

Across the frozen matrix, the artifact pairs all 1,960 normalized prediction rows with 1,960 raw provider receipts and records 27 run manifests. The adapter validates response-reported model identity, while request bodies exclude gold, annotator, and adjudication fields. This supports deterministic replay from an accepted proposal, provider inspection from a receipt, and aggregate reconstruction from the ledger and frozen hashes. Before release, engineers can compare this ledger with the preceding frozen contract and inspect changed dispatch decisions. This comparison turns release review into a contract-level change audit across model and code boundaries.

Fail-closed behavior is configurable only to one of the four non-dispatch terminals; \textsc{dispatch} is rejected as a fallback during controller construction. Persistent JSON or contract failure therefore ends at a recorded bounded outcome. Semantically wrong but contract-valid proposals remain visible in the action and final-path metrics and error analysis. This measurement boundary positions \method{} as the dispatch controller within a larger customer-service system. The versioned assets also define the change surface. A rule edit changes admissibility and triggers FLIP regression; a registry edit changes valid object identifiers and triggers ENTITY regression; a catalog or prompt edit changes semantic selection and triggers MAIN plus the known account-analysis/information-lookup confusion cases. Adapter changes require schema, response-identity, timeout, repair, and fail-closed tests. These mappings keep a policy correction from being evaluated only as a prompt change and preserve the primary-commitment invariant across maintenance releases.
 Deployment observability follows the same contract. A decision record carries a stable request identifier, catalog and rule hashes, primary proposal, validation result, winning rule, object-resolution result, final path, repair or fallback status, and backend-reported identity. Request text and provider receipts remain subject to institutional access controls; a public artifact can retain schemas, hashes, synthetic cases, and normalized predictions. This separation allows an incident to be replayed from an accepted proposal without invoking the hosted backend again, while preserving the raw receipt for provider-level inspection.

\section{Experiments and Evaluation}

\subsection{Research Questions and Evaluation Overview}

We evaluate \method{} as a boundary-aware pre-response routing controller in the customer-service setting of a real financial institution. The evaluation separates semantic selection, deterministic policy enforcement, closed-list object validation, and backend choice. A result is useful only when the final path, safety error, and operational trace agree; aggregate accuracy alone cannot distinguish unsafe dispatch from excessive blocking.

The evaluation is organized around four integration decisions. A complete path must first be compared with matched one-call designs under the same information. A state-aware controller must then react to a changed runtime predicate while request-side evidence remains fixed. Closed-list object validation must be isolated from state enforcement because it controls a different dispatch precondition. Finally, the model backend is replaceable, so the same controller and test cases must be exercised across providers. These decisions yield four research questions.

\noindent\textbf{RQ1---Executable routing correctness.} How accurately does \method{} convert a semantic proposal into a complete executable route compared with matched one-call routing designs?\\
\textbf{RQ2---State-conditioned execution.} Under identical request-side input, does deterministic state enforcement produce the correct state-dependent terminal, and what fails when that enforcement is removed?\\
\textbf{RQ3---Closed-list object contract.} How does closed-list object validation affect object correctness and the availability of otherwise admissible object-sensitive routes?\\
\textbf{RQ4---Backend portability and operational variation.} With the controller and evaluation cases fixed, how stable are routing correctness, safety, interface validity, and operational cost across hosted model backends?

Assessment proceeds through four layers. Final-path exact match evaluates the dispatched action or bounded terminal; path macro-F1 and action accuracy expose class and semantic errors. False dispatch counts a dispatch when the gold path is bounded, while false block counts a bounded output when dispatch is admissible. Paired FLIP accuracy requires both state variants to be correct. Object accuracy, valid JSON, calls, tokens, and latency complete the evidence.

Each question has a distinct decision criterion. RQ1 accepts no label-only shortcut: the action, object when required, and terminal must form the gold final path. RQ2 treats the two members of each FLIP pair as one test and uses A-state to expose the consequence of removing deterministic authorization. RQ3 evaluates object resolution only on ENTITY and uses A-ground to measure both object correctness and lost admissible dispatch. RQ4 freezes the controller and cases while replacing only the backend profile. This mapping keeps state and object mechanisms separate, prevents the backend screen from being interpreted as an architectural ablation, and keeps SELECT evidence separate from the CONFIRM evidence used for the main claims.

For a set of $n$ cases, final-path exact match is $n^{-1}\sum_i \mathbf{1}[\hat{y}_i=y_i]$. Macro-F1 averages F1 over the five dispatch actions and four bounded terminals observed in the benchmark. False dispatch uses only gold non-dispatch cases as its denominator; false block uses only gold \textsc{allow} cases. We report both counts and rates so a low rate cannot conceal a small denominator. FLIP pair accuracy counts a pair only when both members are exact. Exact McNemar tests compare paired correctness vectors. The artifact records Wilson intervals for final-path exact match and FLIP pair accuracy. Statistical tests do not treat the two members of a FLIP pair as independent pair-level observations.

Together, these metrics separate semantic selection, authorization, grounding, and availability, while JSON, identity, call, token, and latency fields characterize the adapter and backend boundary.

\subsection{Benchmark, Systems, and Protocol}

The benchmark contains 600 privacy-preserving customer-service scenarios reconstructed from recurring patterns in confidential requests; raw requests and customer identifiers remain confidential. Three engineers from the institution jointly reviewed the scenarios and current labels for practice relevance. The cases cover five service families and 40 public ETF entities whose exchange codes, names, and source records are documented in the released entity ledger. Runtime states and benchmark policies are synthetic. Labels were reviewed jointly without independent duplication, so inter-annotator agreement was not computed.

Construction begins from recurring task and ambiguity patterns instead of copied utterances. Templates vary the requested service, cross-service cues, object-reference form, and runtime boundary, then express each combination as a privacy-preserving request. Public ETF codes, display names, and exchange sources provide real closed identifiers. The released language is reconstructed; customer, account, transaction, and internal-rule records remain confidential. This design exercises the software interfaces while preserving the institution's data boundary.

The construction ledger retains the scenario cell, bounded history, runtime record, candidate registry, cue evidence, action, object difficulty, canonical object, boundary, final path, and applicable rule identifiers. These fields separate the natural-language request from the executable oracle. In particular, a gold action does not determine a gold terminal without the versioned state and rules, and a dispatchable action does not determine a complete path without a valid canonical object. Public entity records also retain their exchange source and verification date, allowing the closed registry to be checked independently of the synthetic request text.

MAIN, FLIP, and ENTITY are evaluation views over the same versioned schema rather than unrelated tasks. MAIN samples the ordinary composition of semantic, state, and object decisions. FLIP duplicates request-side evidence but changes one state record, making the pair itself the regression unit. ENTITY concentrates reference forms for which the action may be admissible but the object contract still determines whether dispatch can complete. All three views use the same final-path type. This common representation permits one controller and one scoring implementation to expose different failure sources without changing the claimed software output between RQs.

Gold fields remain separate from every model request and response receipt. The evaluator reads the frozen action, boundary, canonical object, and final path only after the controller has emitted its result. It then recomputes denominator membership for false dispatch and false block from the gold boundary rather than from the prediction. This ordering matters for fail-closed systems: a malformed response still contributes its configured bounded outcome, while a refusal on an admissible case remains a false block. Thus, interface failure, conservative fallback, and semantic error are counted as software-visible outcomes rather than filtered before scoring.

Before execution, a read-only audit checks case identifiers, label domains, JSON fields, catalog and rule references, and the 40 canonical entity identifiers. It rejects normalized request overlap among the protected non-FLIP partitions and requires every FLIP identifier to contain exactly two records with identical request-side input, different runtime records, and exactly one \textsc{allow} boundary. The audit prints split statistics and SHA-256 hashes without editing any row. Practitioner review establishes practice relevance; the automated audit protects the structural assumptions used by the paired metrics and run scripts.

The 600 scenarios contain 123 cases for each of advisory, information lookup, account analysis, and protected action, plus 108 support cases. Their gold boundaries comprise 262 \textsc{allow}, 128 \textsc{clarify}, 75 \textsc{defer}, 75 \textsc{handoff}, and 60 \textsc{reject} cases. Object-reference difficulty includes exact names, codes, anaphora, conflicting mentions, underspecification, and object-free requests. The controlled distribution targets contract boundaries and mixed-service language; estimating issue frequency in production traffic requires a separately sampled operational dataset.

The 120-case DEV split is used only for prompt and interface checks. SELECT contains 160 cases (100 MAIN, 40 FLIP, and 20 ENTITY) and compares hosted backends. The disjoint CONFIRM set contains the remaining 320 cases (200 MAIN, 80 FLIP, and 40 ENTITY) and evaluates systems and safeguards. Every FLIP subset contains complete same-request pairs. In 23 CONFIRM cases, more than one non-default rule matches; the frozen numeric priority determines the gold routing outcome. All reported calls use the same frozen prompt, catalog, rules, dataset rows, and deterministic validators.

Table~\ref{tab:splits} makes the data flow explicit. SELECT and CONFIRM partition the 480 non-DEV cases without overlap. Their action and boundary distributions retain coverage of all five actions and all five boundary values; paired records are assigned as complete units. MAIN exercises the ordinary mixture of services, objects, and boundaries. FLIP consists of two records with identical request-side input and one state change. ENTITY concentrates exact-code, exact-name, anaphoric, conflicting, and underspecified references.

\begin{table}[!t]
\caption{Frozen benchmark partitions. M/F/E: MAIN, FLIP, and ENTITY; A/B: dispatchable and bounded cases.}
\label{tab:splits}
\centering
\footnotesize
\setlength{\tabcolsep}{3.2pt}
\renewcommand{\arraystretch}{1.14}
\begin{tabular}{@{}lrrrl@{}}
\toprule
Partition & Cases & M/F/E & A/B & Role \\
\midrule
DEV & 120 & -- & 50/70 & Prompt/interface checks \\
SELECT & 160 & 100/40/20 & 76/84 & Backend screening \\
CONFIRM & 320 & 200/80/40 & 136/184 & RQ1--RQ3 \\
\bottomrule
\end{tabular}
\end{table}

We compare \method{} with \textbf{B1 Direct}, which predicts the final route without an explicit rule program, and \textbf{B2 Direct+Guard}, the principal baseline that receives the complete rules and predicts the route in one call. Both use the same catalog, entities, decoding controls, and output validator. RQ2 uses the offline \textbf{A-state} construction check, which bypasses deterministic enforcement; RQ3 uses \textbf{A-ground}, which disables the closed object resolver. A-state and A-ground are mechanism checks, while B2 is the end-to-end architecture comparator. All evaluated systems terminate at dispatch-boundary records; live institutional services remain outside experimental execution.

B1 tests whether an LLM can infer a complete contract directly from request-side information and state without the explicit rule program. B2 is a stronger prompt-embedded guard: it receives the same catalog, object candidates, versioned state, complete rule list, and output contract as \method{}, cites rule identifiers, and predicts action, object, boundary, and terminal itself. The information is therefore matched; the decisive difference is that B2 predicts the transition, whereas ordinary code computes it in \method{}. All three systems use one model call on a valid first pass and the same format-only repair rule.

The comparisons have distinct roles. B1 omits the rule program and diagnoses the gap between plausible semantic routing and the complete contract. B2 controls information availability and provides the principal architecture comparison. A-state and A-ground reuse accepted proposals and modify one executable mechanism, isolating state and object effects under the same semantic prompt. The six-backend panel fixes controller and cases. Thus, B2 tests architecture, B1 diagnoses the semantic-contract gap, the interventions attribute mechanism effects, and the panel tests portability.

The backend panel includes DeepSeek V4 Pro and Flash, GPT-4o and GPT-5.2, and two Qwen configurations. Exact requested and response-reported identifiers, endpoints, and request options are retained in the artifact rather than expanded in the paper. Temperature is zero, reasoning modes are disabled, concurrency is one per task, and one format-only repair is allowed. We present backend screening before RQ1--RQ3 results because the final matrix fixes DeepSeek V4 Pro as the architecture backend. Paired system differences use exact McNemar tests.

The final experiment matrix predesignates DeepSeek V4 Pro for RQ1--RQ3 and uses the disjoint SELECT partition to document the backend choice. The screening table is presented before the RQ result tables for that reason; it is not pooled with CONFIRM. Requests are executed through provider-compatible chat-completion endpoints. Each profile freezes the model name and accepted response identities, and every case stores the reported identity. Because hosted checkpoints can change behind stable aliases, the requested and response-reported identities define the evaluated service profile and its replication scope.

The matrix contains 1,960 system--case executions before repair calls: 600 for RQ1 (three systems on 200 MAIN cases), 320 for RQ2 (three systems and A-state on 80 FLIP cases), 80 for RQ3 (\method{} and A-ground on 40 ENTITY cases), and 960 for RQ4 (six backends on SELECT-160). RQ1--RQ3 use the same primary backend; RQ4 changes only the backend profile. Hashes freeze the controller, prompt, SELECT file, and CONFIRM file. The receipt and request-field audit described in Section~III verifies the evidence chain for every matrix row.

The evaluation gates correspond to the deployed contracts rather than to one aggregate score. MAIN tests the complete action--object--terminal path, FLIP tests whether a changed predicate changes the path while request evidence is fixed, and ENTITY tests whether closed-list object validation preserves admissible dispatch. False dispatch and false block are sliced by their applicable denominators, while JSON validity, repair, identity, calls, and latency characterize the adapter boundary. Evidence of improvement therefore requires path correctness together with both error types; indiscriminate blocking and label-only success fail this criterion.

The protocol also separates semantic and executable acceptance. First, the adapter must return contract-valid JSON with identifiers drawn from the supplied catalog and registry. Second, the deterministic controller must reproduce the expected boundary and object contract from frozen inputs. Third, the final path is compared with gold, including the exact bounded terminal. Finally, the ledger checks model identity, repair, fallback, call count, token use, and latency. A case that fails an earlier interface remains in the denominator with its configured fail-closed outcome. This ordering matches the deployed control path and prevents interface failures from disappearing during metric aggregation.

\begin{table}[!t]
\caption{Backend screening on SELECT-160. Values are percentages except calls, tokens, and median latency. FD: false dispatch; FB: false block.}
\label{tab:backend}
\centering
\scriptsize
\setlength{\tabcolsep}{2.0pt}
\renewcommand{\arraystretch}{1.12}
\begin{tabular}{@{}lrrrrrr@{}}
\toprule
Backend & Path & FD & FB & JSON & Calls & Tok./ms \\
\midrule
DeepSeek V4 Pro & \textbf{97.5} & 0.0 & 1.3 & 100.0 & 1.00 & 762/1367 \\
DeepSeek V4 Flash & \textbf{97.5} & 0.0 & 2.6 & 99.4 & 1.01 & 767/996 \\
Qwen3.8 Max & \textbf{97.5} & 0.0 & \textbf{0.0} & 100.0 & 1.00 & 737/1930 \\
GPT-4o & 96.9 & 0.0 & 2.6 & 100.0 & 1.00 & 730/1853 \\
Qwen3.7 Plus & 96.2 & 0.0 & 1.3 & 100.0 & 1.00 & 739/2293 \\
GPT-5.2 & 95.6 & 0.0 & 6.6 & 100.0 & 1.00 & 742/2329 \\
\bottomrule
\end{tabular}
\end{table}

\subsection{Confirmatory Contract Evidence}

\noindent\textbf{RQ1---Executable routing correctness.}
Table~\ref{tab:main} reports the matched CONFIRM comparison. On MAIN, \method{} reaches 97.5\% final-path exact match and 96.3\% macro-F1, compared with 41.0\% and 36.2\% for B1. B1 nevertheless obtains 94.0\% action accuracy. Its 118 path errors include 96 false dispatches among the 126 bounded MAIN cases, although all 200 outputs are accepted JSON without fallback. Semantic action accuracy therefore does not establish that the action--object--terminal path is executable. \method{} significantly exceeds B1 ($p < 0.001$).

B2 is the more demanding comparison because it receives the same state, rules, catalog, object candidates, and output contract. It reaches 95.5\% MAIN exact match, and its difference from \method{} is not significant ($p = 0.344$). A well-specified one-call guard is therefore competitive. The architectural distinction is transition ownership: B2 predicts the boundary inside the model output, whereas \method{} computes it from the committed action and versioned state, making the decision directly replayable from the winning rule.

\method{} produces five wrong final paths among the 200 MAIN cases. Four confuse account analysis with information lookup; one support request is proposed as advisory and then clarified because the selected action requires an object. Five additional action errors are masked by a correct non-dispatch terminal. These cases explain why we report both action and final-path measures: a safe terminal can still conceal a semantic defect, while a plausible action can conceal an unsafe path. RQ1 therefore finds that the complete contract, rather than action accuracy alone, is the appropriate acceptance object for pre-response routing.

\begin{table}[!t]
\caption{Evidence for RQ1 and RQ2 on CONFIRM. MAIN and pair values are percentages; safety counts pool 200 MAIN and 80 FLIP cases.}
\label{tab:main}
\centering
\footnotesize
\setlength{\tabcolsep}{3.0pt}
\renewcommand{\arraystretch}{1.15}
\begin{tabular}{@{}lrrrrr@{}}
\toprule
System & MAIN & Macro-F1 & FLIP pair & FD & FB \\
\midrule
\method{} & \textbf{97.5} & \textbf{96.3} & 92.5 & \textbf{0/166} & \textbf{1/114} \\
B1 Direct & 41.0 & 36.2 & 10.0 & 127/166 & 2/114 \\
B2 Direct+Guard & 95.5 & 94.5 & \textbf{100.0} & 1/166 & 2/114 \\
\bottomrule
\end{tabular}
\end{table}

\noindent\textbf{RQ2---State-conditioned execution.}
The 40 FLIP pairs hold request text, history, catalog, and object evidence fixed while changing one runtime predicate. \method{} obtains 37/40 correct pairs (92.5\%); B1 obtains 10.0\%, whereas B2 obtains 100.0\%. B2 therefore leads this paired measure, 40/40 to 37/40. In each of \method{}'s three missed pairs, the proposal confuses account analysis with information lookup. The code-computed boundary remains correct on all 80 members: the bounded member reaches the required terminal, while the admissible member dispatches the wrong service. The residual error is semantic, not a rule-engine failure.

The A-state construction check isolates authorization. Bypassing deterministic enforcement lowers FLIP exact match from 96.2\% to 46.2\%, obtains 0/40 correct pairs, and dispatches all 40 members that require a bounded terminal. This mechanism intervention identifies the consequence of removing code-owned $b(a,z,\mathcal{R})$, while B2 supplies the architecture comparison. RQ2 therefore finds that paired same-request tests expose state-conditioned execution within the frozen equality-predicate policies.

Across MAIN and FLIP, \method{} produces 0 false dispatches among 166 bounded cases and one false block among 114 dispatchable cases. B1 dispatches 127/166 bounded cases; B2 dispatches 1/166. These counts complement pair accuracy: B2 is perfect on the FLIP oracle, while \method{} supplies a transition that can be reproduced from versioned code and the recorded winning rule.

\noindent\textbf{RQ3---Closed-list object contract.}
On ENTITY, full \method{} reaches 100.0\% final-path exact match and 95.0\% object accuracy. Disabling the resolver in the A-ground construction check yields 45.0\% and 42.5\%, and blocks all 22 dispatchable object cases. This mechanism check isolates the object contract's effect on object correctness and admissible dispatch. Unlike state enforcement, it determines whether an admissible object-sensitive request can complete.

The two object mismatches made by full \method{} occur on already bounded \textsc{clarify} cases, so both retain the correct terminal. Four additional account-analysis requests are proposed as information lookup but also remain bounded. The trace therefore preserves diagnostic evidence even when the software-visible path is conservative. By contrast, A-ground converts every admissible object-sensitive route to \textsc{clarify}; its zero false-dispatch count is achieved through complete loss of availability on that subset. RQ3 consequently shows that object correctness and false block must be assessed together.

\begin{table}[!t]
\caption{Controlled construction checks for RQ2 and RQ3. Obj. is object accuracy; FD and FB use their applicable denominators.}
\label{tab:ablation}
\centering
\footnotesize
\setlength{\tabcolsep}{3.2pt}
\renewcommand{\arraystretch}{1.14}
\begin{tabular}{@{}llrrrr@{}}
\toprule
System & Split & Path & Obj. & FD & FB \\
\midrule
\method{} & FLIP & 96.2 & 100.0 & 0/40 & 0/40 \\
A-state & FLIP & 46.2 & 100.0 & 40/40 & 0/40 \\
\method{} & ENTITY & 100.0 & 95.0 & 0/18 & 0/22 \\
A-ground & ENTITY & 45.0 & 42.5 & 0/18 & 22/22 \\
\bottomrule
\end{tabular}
\end{table}

Across the 1,000 CONFIRM executions assigned to RQ1--RQ3, no run requires repair, fallback, or a transport-error outcome. Interface reliability therefore does not explain the measured differences among the systems and interventions.

\subsection{Backend Portability}

\noindent\textbf{RQ4---Backend portability and operational variation.}
Table~\ref{tab:backend} shows that all six backends remain within 1.9 percentage points on final-path exact match. DeepSeek V4 Pro, V4 Flash, and Qwen3.8 Max each reach 97.5\%; GPT-4o reaches 96.9\%, Qwen3.7 Plus reaches 96.2\%, and GPT-5.2 reaches 95.6\%. None produces a false dispatch on SELECT. Qwen3.8 Max produces no false block, while V4 Pro and Qwen3.7 Plus each produce one. V4 Flash is fastest, but one of the 960 backend--case executions remains invalid after bounded repair and reaches the fail-closed fallback, yielding 99.4\% accepted JSON and 1.01 calls per case. The narrow path-score spread indicates contract portability; false blocks, JSON acceptance, and latency distinguish the profiles.

The latency and token columns describe hosted-service operation at the time of access and include network and provider variation. Tokens remain close because all profiles receive the same payload and output contract; latency varies more substantially across endpoints. The persistent failure remains in the denominator as the software-visible fallback result. Across the six profiles, accepted JSON is 99.4--100\%, calls remain 1.00--1.01 per case, and median latency ranges from 996 to 2,329 ms.

The final matrix fixes V4 Pro as a balanced backend. It ties the highest path score, produces zero false dispatch and 100\% accepted JSON, and is faster than the tied Qwen3.8 Max; Flash is faster but has one persistent contract failure and more false blocks. Qwen3.8 Max has the lowest false-block count. These tradeoffs motivate V4 Pro as the fixed RQ1--RQ3 representative and separate architecture effects from backend changes.

Backend errors concentrate in semantic selection and excessive blocking rather than unsafe progress because deterministic enforcement is shared. This is the intended replacement boundary: changing the model can affect which catalog action and object are proposed, but it cannot change rule priority or turn a non-\textsc{allow} decision into dispatch. RQ4 therefore establishes a replacement protocol for the evaluated profiles: rerun the frozen regression suite after changing the proposer. Evaluation across other languages, service catalogs, and later hosted checkpoints remains future work.

\section{Discussion}

\subsection{Engineering Implications and Deployment}

RQ1 changes the acceptance object for customer-service routing. B1's 94.0\% action accuracy coexists with only 41.0\% final-path exact match, so a service label cannot serve as a release gate when software must also select an object and terminal. B2 reaches 95.5\% versus \method{}'s 97.5\%, and the difference is not statistically significant; B2 also wins the FLIP-pair comparison 40/40 to 37/40. The engineering distinction is authority and evidence: the proposal remains untrusted, code owns the executable transition, and the winning rule and provider receipt remain available for replay.

RQ2 and RQ3 separate failure classes hidden by aggregate safety. The A-state and A-ground checks confirm designed consequences: removing state enforcement dispatches all 40 prohibited FLIP members, whereas removing object validation blocks all 22 admissible object-sensitive cases. B2 remains the architecture comparison, while these contract checks isolate state and object effects. State enforcement protects authorization; object validation protects grounding and availability. The three FLIP pairs with a wrong service remain semantic errors.

The boundary localizes maintenance: action confusion belongs to the catalog and proposer, a state error to policy, an unknown identifier to the registry, malformed output to the adapter, and failed-contract dispatch to the controller. RQ4 demonstrates proposer replacement under a frozen controller contract: path accuracy spans 1.9 points while latency, false blocks, and semantic choices differ. Each component change activates its corresponding frozen regression slice. Test selection should also be examined across objectives and distribution shifts instead of one aggregate criterion~\cite{zhang2026empirical}. The next validation step is an external institutional batch followed by privacy-preserving shadow operation.

\subsection{Threats to Validity}

Validity depends on information parity, metric fit, and backend stability. B1, B2, and \method{} share the request, catalog, object candidates, and output schema; B2 also receives the same state and rules as the primary matched control. A-state and A-ground each remove one enforcement function as controlled mechanism tests. Final-path exact match, false dispatch, false block, and FLIP pair accuracy measure the dispatch boundary; response quality and customer outcomes form a separate evaluation layer. Exact McNemar tests and Wilson intervals are retained in the artifact. All 1,960 rows preserve predictions, backend-reported identities, raw receipts, and catalog, rule, and split hashes for replay. Hosted-service changes trigger the frozen regression suite.

\subsection{Scope and Generalizability}

The evaluation is a privacy-preserving, controlled study of routing contracts. Its 600 reconstructed scenarios come from one financial institution and were jointly reviewed by three engineers; customer logs remain confidential. Runtime states and policies are synthetic, while 40 ETF identifiers are public. DEV, SELECT, and CONFIRM are disjoint, and every output is regenerated under frozen hashes; CONFIRM therefore provides internal confirmation within this collection. Separately sampled external data are needed to estimate traffic prevalence and cross-institutional performance.

Transfer is bounded by the closed ontology, equality predicates, single committed object, and hosted backends used in the study. The reported metrics focus on routing correctness and admissibility at the dispatch boundary. End-to-end response quality and live operational outcomes form the next evaluation layer. Future evaluation should extend the contract to range, temporal, and cross-field predicates and composite requests, test an independently reviewed institutional batch, and assess the controller through privacy-preserving shadow deployment.

\section{Conclusion}

We formalize boundary-aware pre-response routing and present \method{}, where a model proposes a route while code owns state authorization and closed-list validation. Our benchmark contains 600 privacy-preserving customer-service scenarios from a real financial institution, jointly reviewed by three engineers. On 200 MAIN cases, \method{} reaches 97.5\% final-path exact match; matched B2 remains competitive at 95.5\%, while B1's 94.0\% action accuracy yields 41.0\% complete-path accuracy. \method{} produces no false dispatch across 166 bounded MAIN and FLIP cases, and receipts support replay of all 1,960 matrix rows. These results support complete-route acceptance and code-owned execution within the studied closed routing setting.

\noindent\textbf{Acknowledgement.} The work described in this paper was substantially supported by Hong Kong Metropolitan University Research Grant (Project Reference No. RD/2026/1.15). This work is also supported in part by the Research Grants Council of Hong Kong and research funds from the City University of Hong Kong (9229029, 9229192, 6000871).

\clearpage
\bibliographystyle{IEEEtran}
\bibliography{references}
\end{document}